\documentclass[12pt]{article}
\usepackage{fullpage}
\usepackage{amssymb, amsmath}
\usepackage{graphicx}
\usepackage{xcolor}
\usepackage{booktabs}
\usepackage{multirow}
\usepackage{algorithm}
\usepackage{setspace}
\usepackage{hyperref}

\newcommand{\argmin}{\operatorname*{arg \ min}}

 \newcommand{\bitem}{\begin{itemize}}
 \newcommand{\eitem}{\end{itemize}}

 \newcommand{\minim}{\operatorname*{minimize}}

\newtheorem{theorem}{Theorem}

\newtheorem{lemma}{Lemma}

\usepackage[figuresright]{rotating}
\usepackage{natbib}

\title{Multiresolution categorical regression for interpretable cell type annotation}
\author
{Aaron J. Molstad\footnote{Correspondence: amolstad@umn.edu}~ and Keshav Motwani$^\dagger$ \\
School of Statistics, University of Minnesota$^*$\\
Department of Biostatistics, University of Washington$^\dagger$}
\date{}
\begin{document}




\label{firstpage}


\maketitle
\begin{abstract}
In many categorical response regression applications, the response categories admit a multiresolution structure. That is, subsets of the response categories may naturally be combined into coarser response categories. In such applications, practitioners are often interested in estimating the resolution at which a predictor affects the response category probabilities. In this article, we propose a method for fitting the multinomial logistic regression model in high dimensions that addresses this problem in a unified and data-driven way. In particular, our method allows practitioners to identify which predictors distinguish between coarse categories but not fine categories, which predictors distinguish between fine categories, and which predictors are irrelevant. For model fitting, we propose a scalable algorithm that can be applied when the coarse categories are defined by either overlapping or nonoverlapping sets of fine categories. Statistical properties of our method reveal that it can take advantage of this multiresolution structure in a way existing estimators cannot. We use our method to model cell type probabilities as a function of a cell's gene expression profile (i.e., cell type annotation). Our fitted model provides novel biological insights which may be useful for future automated and manual cell type annotation methodology.
\end{abstract}
\onehalfspacing

\section{Introduction \label{section intro}}
Multinomial logistic regression is one of the most widely used approaches for modeling the probability mass function of a $K$-category random variable as a function of a $p$-dimensional predictor \citep[Chapter 8]{agresti2003categorical}. This model assumes that the random response category counts for the $i$th subject, $Y_i$, can be modeled 
\begin{equation}\label{eq:multinomiaL_model}
Y_i \sim {\rm Multinomial}\{n_i, \pi_1^*(x_i),\dots, \pi_K^*(x_i)\},  ~~~~ i \in \{1, \dots, n\},
\end{equation}
where $n_i$ is the number of independent trials for the $i$th subject, $x_i = (1, x_{i,2}, \dots, x_{i,p})^\top \in \mathbb{R}^{p}$ is the $p$-dimensional predictor for the $i$th subject, and $\pi_k^*(x_i)$ is the probability of the $k$th category in a single trial for the $i$th subject for $k \in \{1, \dots, K\} =: [K]$. The multinomial logistic regression model assumes the link
\begin{equation}\label{eq:probs}
\pi^*_k(x_i) = \frac{{\rm exp}(x_i^\top \beta_{:,k}^*)}{\sum_{j=1}^K {\rm exp}(x_i^\top \beta_{:,j}^*)}, ~~~~~~ k \in [K],
\end{equation}
where $\beta^* = (\beta_{:,1}^*, \dots, \beta_{:,K}^*) \in \mathbb{R}^{p \times K}$ is an unknown matrix of regression coefficients. For the remainder, let $\beta^*_{j,:} \in \mathbb{R}^K$ be the $j$th row of $\beta^*$, let $\beta_{:,k}^* \in \mathbb{R}^p$ be the $k$th column of $\beta^*$, and let $[a]$ denote the set $\{1, \dots, a\}$ for positive integer $a$.

In this article, we focus on fitting \eqref{eq:probs} in settings where the $K$ response categories can be coarsened at multiple resolutions. By coarsened, we mean that the union of two or more of the $K$ response categories constitutes a coherent (as determined by the application) category.
As a toy example, suppose the set of response categories were $\{$Sweden, Norway, United States, Canada, Brazil, Argentina$\}$. Possible coarsened categories include {Scandinavia} = $\{$Sweden, Norway$\}$, {North America} = $\{$United States, Canada$\}$, {Northern Hemisphere} = $\{$Sweden, Norway, United States, Canada$\}$, and Americas = $\{$United States, Canada, Brazil, Argentina$\}$. 
In applications where the response exhibits this structure, a practitioner is often interested in understanding the resolution at which a predictor affects the response. In the preceding example, a practitioner may consider fitting separate versions of \eqref{eq:probs} with responses $\{${Northern Hemisphere, Southern Hemisphere}$\}$, or $\{${Scandinavia, North America, South America}$\}$, or $\{${Scandinavia, Americas}$\}$, and so on. Each fitted model may provide insights about the relationship between predictor and response which are obscured by the other models. When the dimension of the predictor $p$, number of response categories $K$, and number of coarse categories are large, this becomes a combinatorially difficult modeling task.

Our work is motivated by the task of assigning an individual cell a label corresponding to its biological function, i.e., cell type annotation---a key step in many single-cell RNA-seq data analyses \citep{pasquini2021automated}. Specifically, we model cell type probabilities as a function of a cell's $p$-dimensional gene expression profile measured by single-cell RNA sequencing. One crucial aspect of building such a model is determining which genes' expression distinguishes different cell types. A well-estimated set of discriminative genes will not only yield better classification accuracy, but can provide useful scientific insights and improve model interpretability \citep{dumitrascu2021optimal}. Thus, our goal is to develop a method which performs model estimation and gene selection jointly for the purpose of interpretable cell type annotation. 

\begin{table}[t]
\begin{center}
\caption{Coarse categories in the single-cell dataset from \citet{hao2021integrated}. Indented coarse cell type descriptions indicate that the cell type is a subset of its preceding coarse cell type. Four fine cell types (Eryth, HSPC, ILC, Platelet) do not belong to any coarse category and do not appear in the table. Superscripts are omitted from fine cell types for ease of display. For details, see Figure 2 of \citet{maecker2012standardizing} and references therein.  }\label{tab:coarse_cell_types}
\scalebox{.75}{
\begin{tabular}{ll}
\toprule
\textbf{Coarse cell type description} & \textbf{Fine cell types}\\
\midrule\midrule 
\textbf{B cells} & B intermediate, B memory, B naive, Plasmablast\\
\midrule
\textbf{Monocytes} & CD14 Mono, CD16 Mono\\
\midrule
\multirow{2}{*}{\textbf{T cells}}& CD4 CTL, CD4 Naive, CD4 TCM, CD4 TEM, Treg Memory, Treg Naive,\\
&  CD8 Naive, CD8 TCM, CD8 TEM, dnT, gdT, MAIT\\
\midrule
\hspace{7pt} \textbf{CD4$^+$ T cells} & CD4 CTL CD4 Naive, CD4 TCM, CD4 TEM, Treg Memory, Treg Naive\\
\midrule
\hspace{20pt}  \textbf{CD4$^+$ naive T cells} & CD4 Naive, Treg Naive\\
\midrule
\hspace{20pt}  \textbf{CD4$^+$ memory T cells} & CD4 TCM, CD4 TEM, Treg Memory\\
\midrule
\hspace{7pt}  \textbf{CD8$^+$ T cells} & CD8 Naive, CD8 TCM, CD8 TEM\\
\midrule
\hspace{20pt}  \textbf{CD8$^+$ memory T cells} & CD8 TCM, CD8 TEM\\
\midrule
 \textbf{NK cells} & NK, NK\_CD56bright\\
\midrule
\textbf{Dentritic cells} & ASDC, cDC1, cDC2, pDC\\
\midrule
\hspace{7pt} \textbf{Conventional dentritic cells} & cDC1, cDC2\\
\bottomrule
\end{tabular}
}
\end{center}
\end{table}

Cell type, the categorical response, naturally admits the aforementioned ``multiresolution'' structure. For example, a subset of the cell types in our motivating data analysis include $\{$CD8$^+$ Naive, CD8$^+$ TCM, CD8$^+$ TEM$\}$, where TCM and TEM denote central memory and effector memory T cells, respectively. In this setting, possible coarsened categories of these cell types include
 CD8$^+$ T cells = $\{$CD8$^+$ Naive, CD8$^+$ TCM, CD8$^+$ TEM$\}$ and CD8$^+$ memory T cells = $\{$CD8$^+$ TCM, CD8$^+$ TEM$\}$. A particular gene may be useful for distinguishing CD8$^+$ T cells from the others, but may not be useful for discriminating between the subtypes of CD8$^+$ T cells. Conversely, a gene may distinguish between CD8$^+$ Naive and CD8$^+$ memory T cells, but cannot distinguish between the two subtypes of CD8$^+$ memory T cells. To the best of our knowledge, our method is the first that performs gene selection wherein the effect of a particular gene can be interpreted in such a way. In so doing, our method addresses a recurring theme among the ``grand challenges'' in single-cell data science \citep{lahnemann2020eleven} by accounting for the multiresolution (``varying levels of resolution") of cell types.  In Table \ref{tab:coarse_cell_types}, we provide the set of coarse categories we use in our analysis in Section \ref{sec:real_data}.

The new estimator proposed in this article, loosely speaking, allows practitioners to consider models at all possible response resolutions simultaneously in a unified and data-driven way. 
As we will demonstrate, exploiting the multiresolution structure of the response categories can lead to more interpretable fitted models and better estimates of the mass function of interest, \eqref{eq:multinomiaL_model}, than existing approaches. 

\section{Multiresolution categorical regression}\label{subsec:multires}
\subsection{Parametric constraints}\label{subsec:constraints}
\begin{figure}
\centering
\includegraphics[width=9cm]{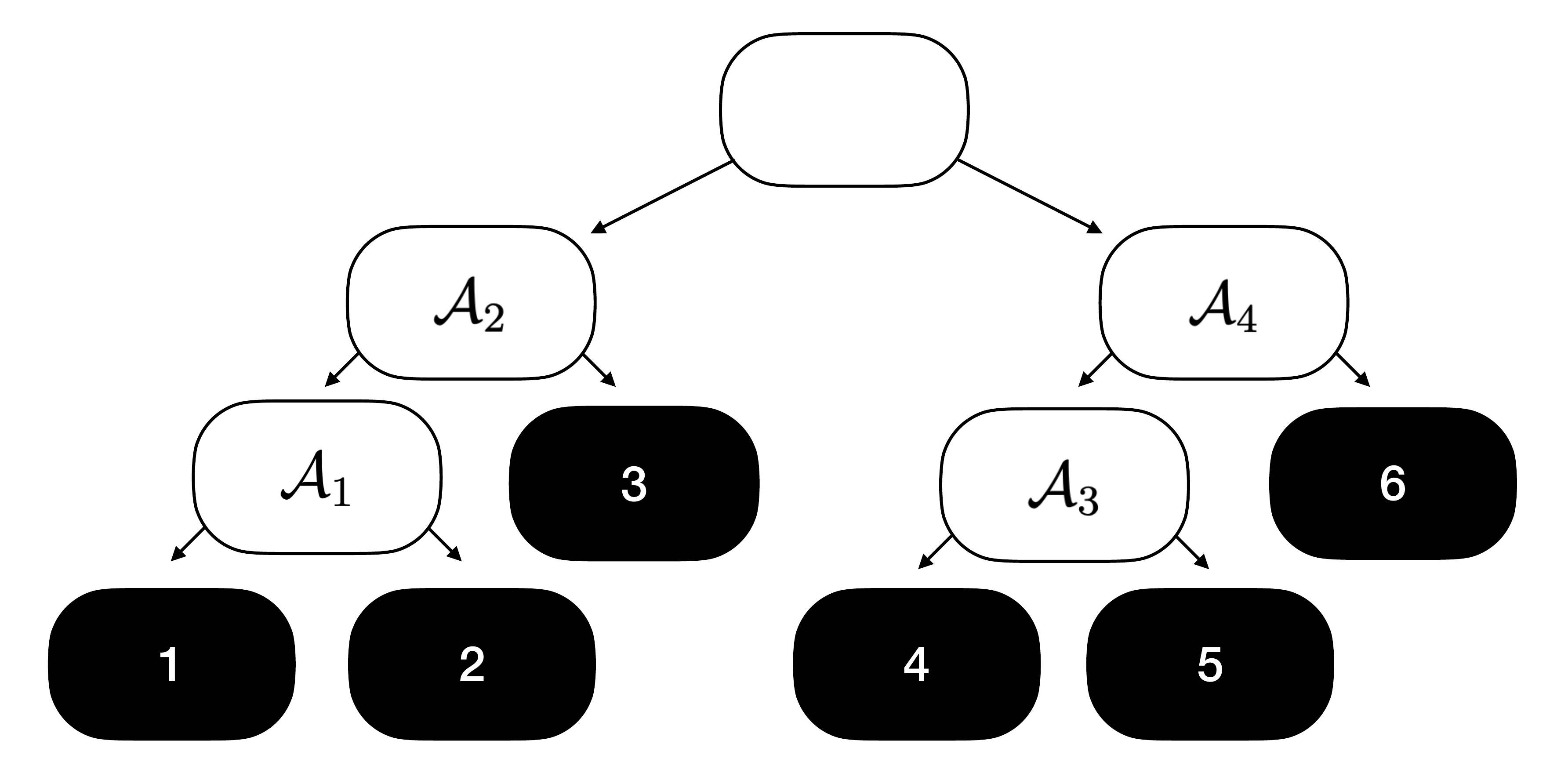}~~~~~~\includegraphics[width=6.5cm]{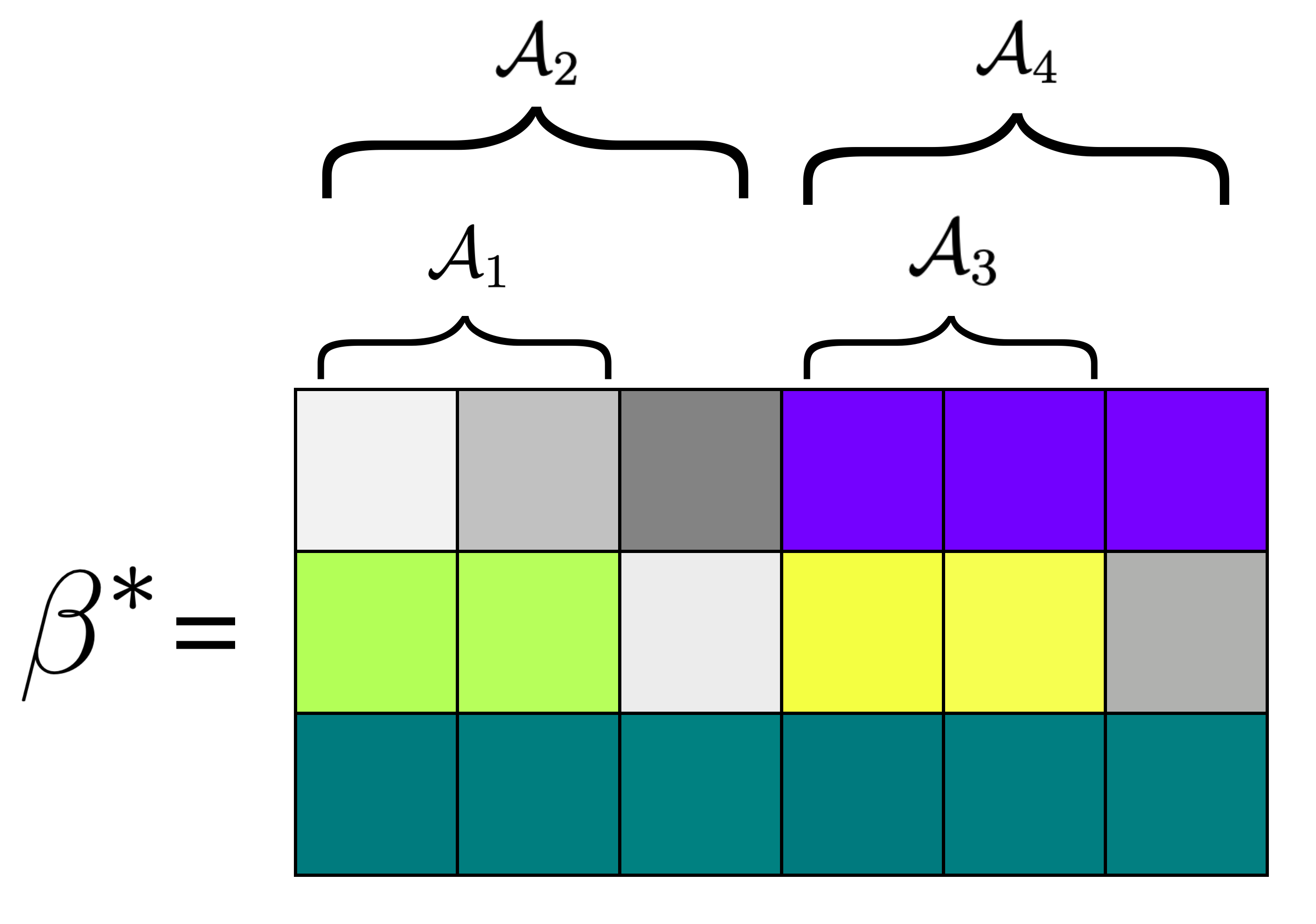}\\
\hspace{50pt}(a)  \hspace{190pt}~~~~~~~~~~ (b) \\
\caption{(Left) A tree depicting the structure of the coarse categories $\mathcal{A}_1, \dots, \mathcal{A}_4$. Note that here, we define $\mathcal{A}_k$ as the set of all fine categories (black leaf nodes) which have $\mathcal{A}_k$ as an ancestor. (Right) A hypothetical regression coefficient submatrix where each cell represents a regression coefficient. Within each row, a distinct color represents a distinct numeric value. For example, the top row has four distinct coefficient values, and the coefficients corresponding to categories in $\mathcal{A}_4$ are all identical. }\label{fig:cases}
\end{figure}

As an example, consider a response variable with (nominal) response category set $\mathcal{K} = \{1, \dots, 6\} =: [6]$. 
 Suppose $\{1, 2\}$ can be coarsened into category $\mathcal{A}_1$, $\{1, 2, 3\}$ into category $\mathcal{A}_2$, $\{4, 5\}$ into category $\mathcal{A}_3$, and $\{4, 5, 6\}$ into $\mathcal{A}_6$. We display the hierarchical structure of these categories in Figure \ref{fig:cases}(a). We refer to the elements of $\mathcal{K}$ as ``fine'' categories to distinguish them from the coarse categories. Each $\pi_k^*$ from \eqref{eq:multinomiaL_model} corresponds to one fine category.   

 To see how one could consider models at various response resolutions simultaneously, suppose all but the $j$th predictor is held fixed. Then, the $j$th predictor can affect the response probabilities at one of multiple possible resolutions: the coarsest resolution, the finest resolution, at a coarse resolution for some categories and fine resolution for others, or not at all. Formally, we say the $j$th predictor does not distinguish between fine categories in $\mathcal{A}_l$ if for any $x \in \mathbb{R}^p$ and $\delta \in \mathbb{R}$, 
\begin{equation}\label{eq:ratios}
\frac{\pi^*_k(x)}{\pi^*_{k'}(x)} = \frac{\pi^*_k(x + \delta e_j)}{\pi^*_{k'}(x + \delta e_j)},
\end{equation}
    for all pairs $(k, k') \in \mathcal{A}_l \times  \mathcal{A}_l$, where $e_j$ is $j$th standard basis vector in $\mathbb{R}^p$.  That is, \eqref{eq:ratios} says that the probability of the response taking category $k$ relative to category $k'$ is not affected by the $j$th predictor.  Under model \eqref{eq:multinomiaL_model}, \eqref{eq:ratios} holds if 
    $\beta^*_{j,k} - \beta^*_{j,k'} = 0$. Thus, restated in terms of $\beta^*$, the $j$th predictor does not distinguish between fine categories in $\mathcal{A}_l$ if $\beta^*_{j,\mathcal{A}_l}\propto 1_{a_l},$
    where $a_l = |\mathcal{A}_l|$, $1_{a_l} = (1, \dots, 1)^\top \in \mathbb{R}^{a_l},$ and  $\beta^*_{j,\mathcal{A}_l} \in \mathbb{R}^{a_l}$ is the subvector of $\beta^*_{j,:}$ consisting of the elements indexed by $\mathcal{A}_l.$

    Continuing with the example with six fine response categories, we provide a hypothetical submatrix of $\beta^*$ in Figure \ref{fig:cases}(b). In this figure, each cell represents a regression coefficient, and each row corresponds to a distinct predictor. Within a row, each distinct color represents a distinct numeric value. For example, the leftmost three coefficients in the top row are distinct, and the rightmost three coefficients are all identical (though distinct from the leftmost three). Based on this structure, the corresponding predictor distinguishes between all fine response categories in $\mathcal{A}_1$ and $\mathcal{A}_2$, but does not distinguish between the fine categories in $\mathcal{A}_4$. By the same logic, the predictor corresponding to the middle row distinguishes categories 3 and 6 from all others, but does not distinguish between fine categories in coarse categories $\mathcal{A}_1$ or $\mathcal{A}_3$. Finally, because all the coefficients in the bottom row are identical, the corresponding predictor does not distinguish between any of the response categories. Hence, if $\beta^*_{j,:} \propto 1_K$, then the $j$th predictor is irrelevant (i.e., do not distinguish between any categories).

    In the next section, we propose a method which allows one to explore these types of blockwise structures along a continuous solution path using convex optimization. The path is controlled by a pair of tuning parameters which can be chosen using cross-validation, i.e., the user need not specify how the predictor affects the response. 

It is important to emphasize that our method does not require that fine categories belong to a single coarse category, nor does it require that coarse categories adhere to a tree-structured hierarchy (as in Figure \ref{fig:cases}). All that our method requires is a collection of sets $\mathcal{A}_l \subset [K]$ for $l \in  [L]$ such that all the categories whose indices belong to $\mathcal{A}_l$ constitute a meaningful (as determined by the application) coarse category.

\subsection{Penalized maximum likelihood estimator}\label{sec:pen_max_lik}
Let $y_1, \dots, y_n$ be $n$ independent realizations from \eqref{eq:multinomiaL_model} based on predictors $x_1, \dots, x_n$, respectively. Specifically, $y_{i} = (y_{i,1}, \dots, y_{i,K})^\top$ where $y_{i,k}$ is the count for the $i$th subject's $k$th category. The (scaled by $1/n$) negative log-likelihood for \eqref{eq:multinomiaL_model}, ignoring constants, is
$$ \mathcal{G}(\beta) = \frac{1}{n}  \sum_{i=1}^n \left[ n_i \log \left\{ \sum_{k=1}^K {\rm exp}(\beta_{:,k}^\top x_i) \right\} - \sum_{k=1}^K y_{i,k} (\beta_{:,k}^\top x_i) \right].$$
Without loss of generality, we assume that $n_i = 1$ for $i \in [n]$ for the remainder of the article. 

To obtain estimates of $\beta^*$ which account for the multiresolution structure of the response, we use a penalized maximum likelihood estimator. Recall that $\mathcal{A}_l \subset [K]$ is the set of indices corresponding to the $l$th coarse category, and define $a_l$ as the cardinality of $\mathcal{A}_l$. We propose the multiresolution penalty $\Omega_{\mathcal{A}}$ given by 
$$  \lambda \hspace{1pt}\Omega_{\mathcal{A}} (\beta) =  \lambda \sum_{j=2}^p \sum_{l = 1}^{L} w_l \left( \min_{c \in \mathbb{R}} \|\beta_{j,\mathcal{A}_l} - c 1_{a_l}\|_2\right)$$
where $\lambda> 0$ is a user-specified tuning parameter, $w_l > 0$ is a user-specified weight for the $l$th coarse category, and $\|\cdot\|_2$ is the Euclidean norm. For sufficiently large values of $\lambda$, this penalty will eventually require that for pair $(j,l)$, $\beta_{j, \mathcal{A}_l} = c 1_{a_l}$ for scalar $c \in \mathbb{R}$. As discussed in Section \ref{subsec:constraints}, $\beta^*_{j, \mathcal{A}_l} = c 1_{a_l}$ would imply that the $j$th predictor is not useful for distinguishing between fine categories within the $l$th coarse category. Note that the outer summation in $\lambda \hspace{1pt}\Omega_{\mathcal{A}} (\beta)$ begins at $j=2$ because we do not penalize the intercept by default. Adjusting our method to also penalize the intercept requires only trivial modifications. 

Intuitively, $\Omega_{\mathcal{A}}$ functions like a group lasso penalty \citep{yuan2006model}, but rather than shrinking all coefficients in a group towards zero, it shrinks all coefficients towards a nearby constant. Because the $c$ minimizing the $(j,l)$th term in the penalty is simply the average of $\beta_{j, \mathcal{A}_l}$, $\overline{\beta_{j,\mathcal{A}_l}} = 1_{a_l}^\top \beta_{j, \mathcal{A}_l}/a_l$, 
if we let $D(a_l) = I_{a_l} - 1_{a_l}1_{a_l}^\top /a_l \in \mathbb{R}^{a_l \times a_l}$ for $l \in [L]$, it follows that
\begin{equation}\Omega_{\mathcal{A}} (\beta) = \lambda \sum_{j=2}^p \sum_{l = 1}^{L} w_l \|\beta_{j,\mathcal{A}_l} - \overline{\beta_{j,\mathcal{A}_l}} 1_{a_l}\|_2 = \sum_{j=2}^p \sum_{l=1}^L w_l\|D(a_l)\beta_{j,\mathcal{A}_l}\|_2.\notag
\end{equation}
Based on this characterization, it is clear that $\Omega_{\mathcal{A}} (\beta)$ is a seminorm and thus convex. For the remainder, we take $w_l = 1$ though other choices of $w_l$ may be useful in certain applications. 

While the penalty $\Omega_{\mathcal{A}}$ can exploit the special structure of $\beta^*$ described in Section \ref{subsec:multires}, it will not perform variable selection (i.e., removal of irrelevant predictors). As mentioned in Section \ref{subsec:multires}, for a predictor to be irrelevant to \eqref{eq:multinomiaL_model}, it must be that all entries of the corresponding row of $\beta^*$ are equivalent. Thus, to achieve variable selection, a natural approach is to use a group lasso penalty on the rows of $\beta$. For large values of the tuning parameter corresponding to this penalty, this will require entire rows of $\beta$ to be equal to zero. Beyond variable selection, the group lasso penalty also implicitly imposes the identifiability condition that each penalized row of $\beta$ sum to zero \citep{Molstad2021}.  
With the multiresolution penalty $\Omega_{\mathcal{A}}$ in hand, we propose to estimate $\beta^*$ using
\begin{equation}\label{eq:estimator}
\argmin_{\beta \in \mathbb{R}^{p \times K}} \left\{ \mathcal{G}(\beta) + \gamma \sum_{j=2}^p \|\beta_{j,:}\|_2 + \lambda \Omega_{\mathcal{A}}(\beta)\right\}
\end{equation}
where $\gamma > 0$ is an additional tuning parameter controlling the contribution of the group lasso penalty.  For the remainder of the article, we will refer to the estimator defined in \eqref{eq:estimator} as \texttt{mrMLR}, the \underline{m}ulti\underline{r}esolution \underline{m}ultinomial \underline{l}ogistic \underline{r}egression estimator.

\subsection{Existing methods}\label{sec:related_methods}

An enormous number of methods exist for cell type annotation, e.g., see the survey of \citet{pasquini2021automated} and references therein. The methods most closely related to our own---categorized ``supervised methods'' in \citet{pasquini2021automated}---are often based on nonparametric classifiers like random forests, $k$-nearest neighbors, and neural networks.  These methods are not model-based and thus, fitted models are often difficult to interpret. In addition, these methods often fail to exploit the multiresolution structure of cell types. Methods that do take this structure into account often rely on fitting many conditional models \citep{de2019chetah,bernstein2021cello}---as opposed to directly modeling \eqref{eq:multinomiaL_model}---which further complicates variable selection and predictor effect interpretation. 

Many penalized likelihood-based methods exist for fitting the multinomial logistic regression model with high-dimensional predictors. For example, \citet{zhu2004classification} use a penalty on the squared Frobenius norm of the regression coefficient matrix, whereas \citet{vincent2014sparse} use a sparse group lasso penalty on the rows of $\beta$. Neither of these methods could exploit the special structure of the response discussed in Section 1. Other estimators assume $\beta^*$ has low rank \citep{yee2003reduced,powers2018nuclear}. 
Reduced-rank estimators of $\beta^*$ do not make explicit use of known multiresolution structures, nor can they easily be modified to perform variable selection.

Other work related to our own is that of \citet{price2019automatic} and \citet{nibbering2022multiclass}, both of whom fit \eqref{eq:multinomiaL_model} with a penalty on the Euclidean norm of $\beta_{:,k} - \beta_{:,k'}$ for all $k \neq k'$. This encourages estimates of $\beta^*$ such that their $k$th and $(k')$th columns are identical for some $k \neq k'$. This approach thus ``combines'' response categories as their estimated probabilities would always be identical if $\beta_{:,k}^* = \beta_{:,k'}^*$ (i.e., one could never predict category $k$ over category $k'$ and vice versa). 
For cell type annotation, this is not practically useful. 

In Web Appendices G.1 and G.2, we contrast our work to the related methods of \citet{motwani2021binned} and \citet{Molstad2021}. In G.3, we situate our estimator among existing methods for classification that exploit a known hierarchy between response categories, and among methods for effect aggregation in regression \citep{yan2021rare}.

\section{Computation}\label{sec:Computation}
\subsection{Algorithm overview}
The optimization problem to compute our estimator, \texttt{mrMLR}, is convex. Moreover, the multinomial negative log-likelihood is differentiable and has Lipschitz continuous gradient. We can thus apply modern first-order methods---namely, variants of the proximal gradient descent algorithm---to compute \texttt{mrMLR}. 
The proximal gradient descent algorithm is an iterative procedure that defines its $(t+1)$th iterate as the minimizer of a penalized quadratic approximation to the objective function constructed at the $(t)$th iterate \citep[Section 3.1]{polson2015proximal}. Given step size $\tau > 0$, this algorithm defines the $(t+1)$th iterate of $\beta$, $\beta^{(t+1)}$, as 
\begin{align}
\argmin_{\beta \in \mathbb{R}^{p \times K}}\left\{\frac{1}{2\tau} \left\Vert\beta - \{\beta^{(t)} - \tau \nabla \mathcal{G}(\beta^{(t)})\}\right\Vert_F^2 + \sum_{j=2}^p \left( \gamma \|\beta_{j,:}\|_2 + \lambda \sum_{l=1}^L w_l \|D(a_l)\beta_{j,\mathcal{A}_l} \|_2\right)\right\}.\label{eq:prox_grad_iterate}
\end{align} 
For sufficiently small and fixed $\tau > 0$, the sequence of iterates generated by \eqref{eq:prox_grad_iterate} will converge to a solution to \eqref{eq:estimator} as $t \to \infty$ \citep[Section 3.1]{polson2015proximal}. We present an accelerated version of this algorithm, which we implement, in Web Appendix C. Briefly, the accelerated version extrapolates a search point based on previous iterates and performs a backtracking line search to select the step size $\tau$ \citep[Section 4]{beck2009fast}. 
For simplicity, we discuss the algorithm based on iterate \eqref{eq:prox_grad_iterate}.

The crux of this algorithm is computing \eqref{eq:prox_grad_iterate} efficiently. 
Note that the solution to \eqref{eq:prox_grad_iterate} can be obtained row-by-row of $\beta$, so we focus on computing the $j$th row. For $j = 1$, $\beta^{(t+1)}_{1,:}$ is equal to the first row of $\beta^{(t)} - \tau \nabla \mathcal{G}(\beta^{(t)})$. 
For $j \geq 2$, $\beta^{(t+1)}_{j,:}$ can be expressed 
\begin{align} 
 \argmin_{\nu \in \mathbb{R}^{K}} \left\{ \frac{1}{2}\|\nu - \eta\|_2^2 + \tilde{\gamma}\|\nu\|_2 + \tilde{\lambda} \sum_{l=1}^L w_l\|D(a_l)\nu_{\mathcal{A}_l}\|_2\right\},\label{eq:main_prox}
\end{align}
where $\tilde{\lambda} = \tau \lambda$, $\tilde{\gamma} = \tau \gamma$, $\eta \in \mathbb{R}^K$ is the $j$th row of $\beta^{(t)} - \tau \nabla \mathcal{G}(\beta^{(t)})$, and $\nu_{\mathcal{A}_l}\in \mathbb{R}^{a_l}$ is the subvector of $\nu$ whose components are indexed by $\mathcal{A}_l$ for $l \in [L].$  Hence, computing each iterate \eqref{eq:prox_grad_iterate} requires solving \eqref{eq:main_prox}. For that, we have the following lemma.
\begin{lemma}\label{lemma1} 
The solution to \eqref{eq:main_prox}, denoted $\hat{\nu}_{\tilde{\gamma}, \tilde{\lambda}}$, is given by 
\begin{equation}\label{eq:hatnu}
\hat{\nu}_{\tilde{\gamma}, \tilde{\lambda}} =
\left\{\begin{array}{cl}
0 & : \|\hat{\nu}_{0, \tilde\lambda}\|_2 \leq \tilde{\gamma}\\
(1 - \tilde{\gamma}/\|\hat{\nu}_{0, \tilde\lambda}\|_2) \hat{\nu}_{0, \tilde\lambda} & :\|\hat{\nu}_{0, \tilde\lambda}\|_2 > \tilde{\gamma}
\end{array}\right.,
\end{equation}
where $\hat{\nu}_{0, \tilde\lambda}$ is defined as the solution to \eqref{eq:main_prox} with $\tilde\gamma = 0$, i.e., 
\begin{equation} \label{eq:sub_prox}
\hat{\nu}_{0, \tilde\lambda} = \argmin_{\nu \in \mathbb{R}^{K}}\left\{\frac{1}{2}\|\nu - \eta\|_2^2 + \tilde{\lambda} \sum_{l=1}^L w_l \|D(a_l)\nu_{\mathcal{A}_l}\|_2\right\}.
\end{equation}
\end{lemma}
Based on the result of Lemma \ref{lemma1}---a proof of which can be found in Web Appendix A---we see that if we compute \eqref{eq:sub_prox}, we can immediately obtain \eqref{eq:main_prox} based on the equality \eqref{eq:hatnu}. Of course, \eqref{eq:sub_prox} depends on the $\mathcal{A}_l$. If the $\mathcal{A}_l$ are nonoverlapping (their intersection is empty), \eqref{eq:sub_prox} can be solved in closed form. If any of the $\mathcal{A}_l$ overlap, we must use an iterative procedure to compute \eqref{eq:sub_prox}. We will discuss these cases separately. 

Without loss of generality, we proceed as if every element of $[K]$ belongs to at least one $\mathcal{A}_l$. For any $k$ such that $k \not\in \cup_{l=1}^L \mathcal{A}_l$, the $k$th component of $\hat{\nu}_{0, \tilde\lambda}$ is the $k$th component of $\eta$.

\subsection{Nonoverlapping coarse categories}\label{subsec:nonoverlap_comp}
In the situation that the $\mathcal{A}_l$ are nonoverlapping, we have the following result which provides a closed form for \eqref{eq:sub_prox}. We prove this result in Web Appendix A. 

\begin{theorem}\label{thm:prox_omega}
Suppose the $\mathcal{A}_l$ are nonoverlapping. Let $[\hat\nu_{0, \tilde\lambda}]_{\mathcal{A}_l} \in \mathbb{R}^{a_l}$ (resp. $\eta_{\mathcal{A}_l}$) denote the subvector of $\hat\nu_{0, \tilde\lambda}$ (resp. $\eta$) whose components are indexed by $\mathcal{A}_l$ for $l \in [L]$. It follows that the solution to \eqref{eq:sub_prox}, denoted $\hat{\nu}_{0, \tilde\lambda},$ is defined subvector-by-subvector with
  $$[\hat\nu_{0, \tilde\lambda}]_{\mathcal{A}_l} = \left\{\begin{array}{cl}
  (\eta_{\mathcal{A}_l}^\top 1_{a_l}/{a_l}) 1_{a_l} &:\|D(a_l) \eta_{\mathcal{A}_l}\|_2\hspace{2pt} \leq w_l \tilde\lambda\\
  \big(1 - \frac{w_l\tilde\lambda}{\|D(a_l)\eta_{\mathcal{A}_l}\|_2} \big) \eta_{\mathcal{A}_l} + \frac{w_l\tilde\lambda}{\|D(a_l)\eta_{\mathcal{A}_l}\|_2} (\eta_{\mathcal{A}_l}^\top 1_{a_l}/{a_l}) 1_{a_l} &:\|D(a_l) \eta_{\mathcal{A}_l}\|_2\hspace{2pt} > w_l \tilde\lambda
  \end{array} \right.,~~~ l \in [L].$$

\end{theorem}
Based on Theorem \ref{thm:prox_omega}, we see that if $\lambda$ is sufficiently large, $[\hat\nu_{0, \tilde\lambda}]_{\mathcal{A}_l}$ will simply be the average of $\eta_{\mathcal{A}_l}$ times the $a_l$-dimensional vector of ones. When $\lambda$ is sufficiently small, $[\hat\nu_{0, \tilde\lambda}]_{\mathcal{A}_l}$ will be a convex combination of $\eta_{\mathcal{A}_l}$ and the average of $\eta_{\mathcal{A}_l}$ times the $a_l$-dimensional vector of ones.  

\subsection{Overlapping coarse categories}\label{subsec:overlap_comp}

When the $\mathcal{A}_l$ overlap, there is no closed form solution for \eqref{eq:sub_prox} in general. Instead, we rely on an efficient iterative algorithm to solve the dual problem of \eqref{eq:sub_prox}. The dual problem for \eqref{eq:sub_prox} is 
\begin{equation}\label{eq:dual_problem}
\minim_{\zeta \in \mathbb{R}^{K \times L}} ~~\|\eta  - \sum_{l=1}^L M(\mathcal{A}_l) \zeta_{:,l}\|_2^2, ~~ \text{subject to} ~~~\|\zeta_{:,l}\|_2\hspace{3pt}\leq\hspace{1pt} w_l \tilde\lambda,~~ \zeta_{k,l} = 0 ~~\text{for } k \not\in \mathcal{A}_l, ~~ l \in [L],
\end{equation}
where $M(\mathcal{A}_l) \in \mathbb{R}^{K \times K}$ is a matrix such that $[M(\mathcal{A}_l)]_{\mathcal{A}_l, \mathcal{A}_l} = D(a_l) \in \mathbb{R}^{a_l \times a_l}$ for each $l \in [L]$, and $[M(\mathcal{A}_l)]_{j,k} = 0$ for $(j,k) \not\in \mathcal{A}_l \times \mathcal{A}_l$. Given a minimizer of \eqref{eq:dual_problem}, $\hat\zeta$, the solution to \eqref{eq:sub_prox} is $\eta - \sum_{l=1}^L M(\mathcal{A}_l) \hat{\zeta}_{:,l}$.

To solve \eqref{eq:dual_problem} we use a blockwise coordinate descent algorithm (Algorithm 2 of Web Appendix C) inspired by \citet{yan2017hierarchical}, who studied variations of the overlapping group lasso. This algorithm cycles through each $l \in[L]$, defining the $(r+1)$th iterate of $\zeta_{:,l}$, $\zeta_{:,l}^{(r+1)}$, as the minimizer of \eqref{eq:dual_problem} with respect to $\zeta_{:,l}$ with $\zeta_{:,1}, \dots, \zeta_{:,l-1}$ fixed at their $(r+1)$th iterates and $\zeta_{:,l+1}, \dots, 
\zeta_{:,L}$ fixed at their $r$th iterates. Focusing on the update for $\zeta_{:,l}^{(r+1)}$, we must solve
\begin{equation}\label{eq:BCD_Problem}
\minim_{\zeta_{:,l} \in \mathbb{R}^{K}} ~~ \|\tilde\eta - M(\mathcal{A}_l) \zeta_{:,l}\|_2^2, ~~ \text{subject to} ~~~\|\zeta_{:,l}\|_2\hspace{3pt}\leq\hspace{1pt} w_l \tilde\lambda,~~ \zeta_{k,l} = 0 ~~\text{for } k \not\in \mathcal{A}_l,
\end{equation}
where $\tilde\eta = \eta - \sum_{j=1}^{l-1} M(\mathcal{A}_l) \zeta_{:,j}^{(r+1)} - \sum_{j=l+1}^{L} M(\mathcal{A}_l) \zeta_{:,j}^{(r)}.$ 
A minimizer of \eqref{eq:BCD_Problem} is $$\zeta_{\mathcal{A}_l,l}^{(r+1)} =  D(a_l) \tilde\eta_{\mathcal{A}_l} \min(1, w_l \tilde{\lambda}/\| D(a_l) \tilde\eta_{\mathcal{A}_l} \|_2), \text{
with $\zeta_{k,l}^{(r+1)} = 0$ for each $k \not\in \mathcal{A}_l$}.$$ 
Hence, we can quickly cycle through the $\zeta_{:,l}$ in order to solve \eqref{eq:dual_problem}. Each optimization problem is only $K$-dimensional and we need only solve \eqref{eq:sub_prox} for $j \in \{2, \dots, p\}$ such that $\|[\beta^{(t)} - \tau \nabla \mathcal{G}(\beta^{(t)})]_{j,:}\|_2 > \tilde\gamma$, as it can be shown that if $\|[\beta^{(t)} - \tau \nabla \mathcal{G}(\beta^{(t)})]_{j,:}\|_2 \leq \tilde\gamma$, then the corresponding $\hat\nu_{\tilde\gamma, \tilde\lambda} = 0$. Further refinements of the algorithm for solving \eqref{eq:sub_prox} with overlapping $\mathcal{A}_l$ are discussed in  Web Appendix C. 

\section{Statistical properties}\label{sec:stat_properties}
In this section, we establish an asymptotic error bound for our estimator. To simplify matters, we treat the $x_i$ as nonrandom and assume that the intercept is omitted (i.e., the first entries of the $x_i$ need not be one). Defining the norm $\|\beta\|_{1,2} = \sum_{j} \|\beta_{j,:}\|_2$, we study the estimator $\widehat\beta = \argmin_{\beta \in \mathbb{R}^{p \times J}} \{\mathcal{G}(\beta) + \gamma\|\beta\|_{1,2} + \lambda \sum_{j=1}^p \sum_{l=1}^L \|D(a_l)\beta_{j,\mathcal{A}_l}\|_2\}$ for the remainder of this section. A proof of the ensuing results, which follow from arguments similar to those employed in \citet{Molstad2021}, can be found in Web Appendix B.

Since for any $\beta^* \in \mathbb{R}^{p \times K}$, $\beta^*_d = \beta^* - d 1_K^\top$ for any $d \in \mathbb{R}^p$ will lead to the same probabilities \eqref{eq:probs} for all $x \in \mathbb{R}^p$, we must first define the version of the regression coefficient matrix from \eqref{eq:probs} for which our estimator will be consistent. For a given $\beta^*$ (i.e., any matrix which leads to the population level probabilities for all $x \in \mathbb{R}^p$), our estimator is consistent for $\beta^{\dagger} = \beta^{*} - (\beta^{*} 1_K)1_K^\top/K$, i.e., the version of $\beta^{*}$ with row-wise average zero. 
See Web Appendix B for an explanation.

Next, we discuss the conditions and assumptions needed for our theoretical results. First, we require that $X = (x_1, \dots, x_n)^\top \in \mathbb{R}^{n \times p}$ is appropriately scaled column-wise.
\begin{itemize}
\item[] \textbf{C1.} The predictors are scaled so that $\|X_{:,j}\|_2^2 \leq n$ for all $j \in [p]$.
\end{itemize}
We will also require an assumption on the data generating process. 
\begin{itemize}
\item[] \textbf{A1.} The response $y_i$ is a realization from the $n_i = 1$ multinomial logistic regression model \eqref{eq:multinomiaL_model} with probabilities as defined in \eqref{eq:probs} for each $i \in [n].$
\end{itemize}

Let $\mathcal{S} \subset [p]$ be the set of predictors which are relevant (i.e., $\beta^{\dagger}_{j, :} \neq 0$ for $j \in \mathcal{S}$) and let $\mathcal{S}^c = [p] \setminus \mathcal{S}$. By definition of $\beta^\dagger$, the $k$th predictor is irrelevant if $\beta^\dagger_{k,:} = 0.$ Similarly, for each $\mathcal{A}_l$, define $\mathcal{S}_l = \{k \in [p]: \beta^\dagger_{k,\mathcal{A}_l} \neq c 1_{a_l} \text{ for any } c \in \mathbb{R}\}$ as the set of predictors that distinguish between fine categories belonging to the $l$th coarse category. Naturally, $\mathcal{S}_l^c = [p] \setminus \mathcal{S}_l$ is the set of predictors which do not distinguish between the fine categories in coarse category $l$ (i.e., if $k \in \mathcal{S}_l^c$, then $\beta^\dagger_{k,\mathcal{A}_l} = c 1_{a_l}$ for some $c \in \mathbb{R}$). Define the collection of sets $\widetilde{\mathcal{S}} = \{\mathcal{S}, \mathcal{S}_1, \dots, \mathcal{S}_L\}.$

Our second assumption, which depends on $\mathcal{C}(\widetilde{\mathcal{S}}, \phi)$---a restricted subset of $\mathbb{R}^{p \times K}$ defined in Web Appendix B---is a restricted eigenvalue-type condition. 
\begin{itemize}
\item[] \textbf{A2.} For each $(\phi_1, \phi_2) \in (1, \infty) \times (0, \infty)$, there exists a constant $k$ such that $\kappa(\widetilde{\mathcal{S}}, \phi) \geq k > 0$ where 
$$ \kappa(\widetilde{\mathcal{S}}, \phi) = \hspace{-5pt}\inf_{\Delta \in \mathcal{C}(\widetilde{\mathcal{S}}, \phi)\setminus \{0\}} \hspace{-5pt}\frac{{\rm vec}(\Delta)^\top \nabla^2 \mathcal{G}(\beta^\dagger) {\rm vec}(\Delta)}{\|\Delta\|_F^2}.$$
\end{itemize}
The assumption \textbf{A2} is analogous to the standard restricted eigenvalue condition in linear regression where $\mathcal{G}$ is squared error loss.

Lastly, we define the subspace compatibility constant \citep{negahban2012unified} as
$\Psi_{\mathcal{A}}(\widetilde{\mathcal{S}}) = \sup_{\Delta \in \mathbb{R}^{p \times K}, \|\Delta\|_F = 1} \sum_{l=1}^L \sum_{k \in \mathcal{S}_l} \|D(a_l)\Delta_{k, \mathcal{A}_l}\|_2.$
The compatibility constant, $\Psi_{\mathcal{A}}(\widetilde{\mathcal{S}})$, quantifies the coherence between the penalty $\Omega_{\mathcal{A}}$ and the structure of $\beta^\dagger$. If few predictors distinguish fine categories within most coarse categories, $\Psi_{\mathcal{A}}$ will be small. Conversely, if many predictors distinguish fine categories within many coarse categories, the penalty and the structure of $\beta^\dagger$ do not cohere, so $\Psi_{\mathcal{A}}$ will be large. More concretely, for any set of coarse categories one can show that 
$\Psi_{\mathcal{A}}(\widetilde{\mathcal{S}}) \leq \sum_{l=1}^L \sqrt{|\mathcal{S}_l| | \mathcal{A}_l|}$.  We are finally ready to state our main result. 
\vspace{-10pt}

\begin{theorem}\label{thm:main_thm}
Let $\alpha \in (0,1),$  $\phi_1 > 1,$ and $\phi_2 > 0$ be fixed constants. Suppose \textbf{C1}, \textbf{A1}, and \textbf{A2} hold.  Define ${\rho}_{1} = 3(\phi_1 + 1)$, $\rho_{2} = 3\phi_1 \phi_2$, $s^* = \max \{\sqrt{|\mathcal{S}|}, \Psi_{\mathcal{A}}(\tilde{\mathcal{S}})\}$, and $c_n = \max_{i\in[n]}\|x_i\|_2$. If $\gamma = \phi_1[(K/4n)^{1/2} + \{\log(p/\alpha)/n\}^{1/2}]$, $\lambda = \phi_2 \gamma$, and $c_n  s^*  \gamma\to 0$ as $n \to \infty$, then for $n$ sufficiently large,
$$ \|\widehat\beta - \beta^\dagger\|_F \hspace{2pt} \leq \left(\frac{\rho_1 \sqrt{|\mathcal{S}|} + \rho_2 \Psi_{\mathcal{A}}(\widetilde{\mathcal{S}})}{\kappa(\widetilde{\mathcal{S}}, \phi)} \right) \left( \sqrt{\frac{K}{4n}} + \sqrt{\frac{\log(p/\alpha)}{n}}\right)$$
with probability at least $1 -\alpha$.
\end{theorem}
Of course, $|\mathcal{S}|$ is the number of relevant predictors, whereas the magnitude of $\Psi_{\mathcal{A}}(\widetilde{\mathcal{S}})$ is controlled by the cardinality of the $\mathcal{S}_l$ and $\mathcal{A}_l$. Our bound thus agrees with intuition: if many predictors are important for distinguishing between fine categories within many coarse categories, our penalty $\Omega_{\mathcal{A}}$ will hurt estimation by inappropriately biasing our estimator. Conversely, if many of the $\mathcal{S}_l$ have few elements, the bound can be improved by taking $\phi_2$ large, which, loosely speaking, has the effect of reducing the volume of $\mathcal{C}(\widetilde{\mathcal{S}}, \phi)$, and thus inflating $\kappa(\widetilde{\mathcal{S}}, \phi).$

\section{Simulation studies}\label{sec:sims}
To demonstrate the performance of our method, we compare it to multiple competitors under various data generating models and predictor dimensions. For 100 independent replications in each setting, we generate 500 training and validation samples, and $10^4$ testing samples. For each sample, we first generate $x$, a realization of ${\rm N}_p(0, \Sigma)$ where $\Sigma_{j,k} = 0.7^{|j-k|}$ for $(j,k) \in [p] \times [p].$  Then, we generate the response, a $K=12$ category single-trial multinomial random variable with probabilities
$\pi^*_k(x) = {\rm exp}(x^\top \beta^*_{:,k})/\{\sum_{j=1}^K {\rm exp}(x^\top \beta^*_{:,j})\}$ for $k \in [K].$  

We set $\mathcal{A}_1 = \{1, 2, 3\}$, $\mathcal{A}_2 = \{4, 5, 6\}$, $\mathcal{A}_3 = \{7, 8, 9\},$ and $\mathcal{A}_4 = \{10, 11, 12\}$. We generate $\beta^*$ independently in each replication according to one of six models, Models 1--6. For each model, we first randomly select 18 predictors to be relevant, then randomly select $s$ of those to be useful for distinguishing between coarse categories: the remaining $18-s$ are useful for distinguishing between all fine categories. All other $p-18$ predictors are irrelevant. Nonzero values of $\beta^*$ are drawn independently ${\rm N}(0, 5)$.  For $j \in \{1, \dots, 6\}$, Model $j$ sets $s = 3(j-1)$. Under Model 1, all predictors are useful for discriminating between fine categories. Under Model 6, only 3 of the 18 predictors are useful for discriminating between all fine categories. Models 2--5 are intermediate to Models 1 and 6 in terms of the number of the 18 relevant predictors which are useful for distinguishing all fine categories. We provide pseudocode for constructing $\beta^*$ in Web Appendix E.

We compare our estimator \texttt{mrMLR}, to multiple competitors. The first competitor, which we call \texttt{Group}, is a special case of \eqref{eq:estimator} with $\lambda = 0$. By fixing $\lambda = 0$, \texttt{Group} can perform variable selection, but fails to exploit the multiresolution structure of the response categories. The second competitor, \texttt{L1}, is simply
$\argmin_{\beta \in \mathbb{R}^{p \times K}}\{ \mathcal{G}(\beta) + \gamma \sum_{j=2}^p \sum_{k=1}^K |\beta_{j,k}| \}.$
Unlike \texttt{mrMLR} and \texttt{Group}, \texttt{L1} does not encourage variable selection directly. However, we include it because it is arguably more flexible than \texttt{Group}, and thus may be able to account for multiresolution structure of the response categories implicitly.

We also compare to a two-step approximation to \texttt{mrMLR}, \texttt{Approx}, which can take advantage of the multiresolution structure using multiple conditional models.  We first describe this method in the context of nonoverlapping $\mathcal{A}_l.$
In the first step, we construct surrogate responses $\tilde{y}_i \in \{0,1\}^L$ where $\tilde{y}_{i,l} = \sum_{j \in \mathcal{A}_l} \mathbf{1}(y_{i,j} = 1)$ for $i \in [n].$ Then, we fit \eqref{eq:estimator} with $\lambda = 0$ to the regression of the $\tilde{y}_i$, an $L$-category response, on the $x_i$ and select tuning parameters by cross-validation. 
In the second step, we fit $L$ separate conditional logistic regression models where the $l$th model has $a_l$ response categories and uses only data for subjects where $\sum_{j \in \mathcal{A}_l} \mathbf{1}(y_{i,j} = 1) = 1$ (i.e., the $l$th model conditions on $\tilde{y}_{i,l} = 1$). Just as in the first step, each model is fit using \eqref{eq:estimator} with $\lambda = 0$ and tuning parameters chosen by cross-validation. With this set of fitted models, the method \texttt{Approx} then predicts $\pi^*_j(x_i)$ using $\sum_{l=1}^L \widehat{{\rm Pr}}_1(\tilde{Y}_{i,l} = 1 \mid x_i) \widehat{{\rm Pr}}_2(Y_{i,j} = 1 \mid x_i, \tilde{Y}_{i,l} = 1)$ where $\widehat{{\rm Pr}}_1$ is an estimated probability from the model fit in the first step and $\widehat{{\rm Pr}}_2$ is the estimated conditional probability from the model fit in the second step.  Unlike \texttt{Group} and \texttt{L1}, this method both performs variable selection and can take advantage of the special structure of the responses.

\begin{figure}[t]
\centering
\includegraphics[width=\textwidth]{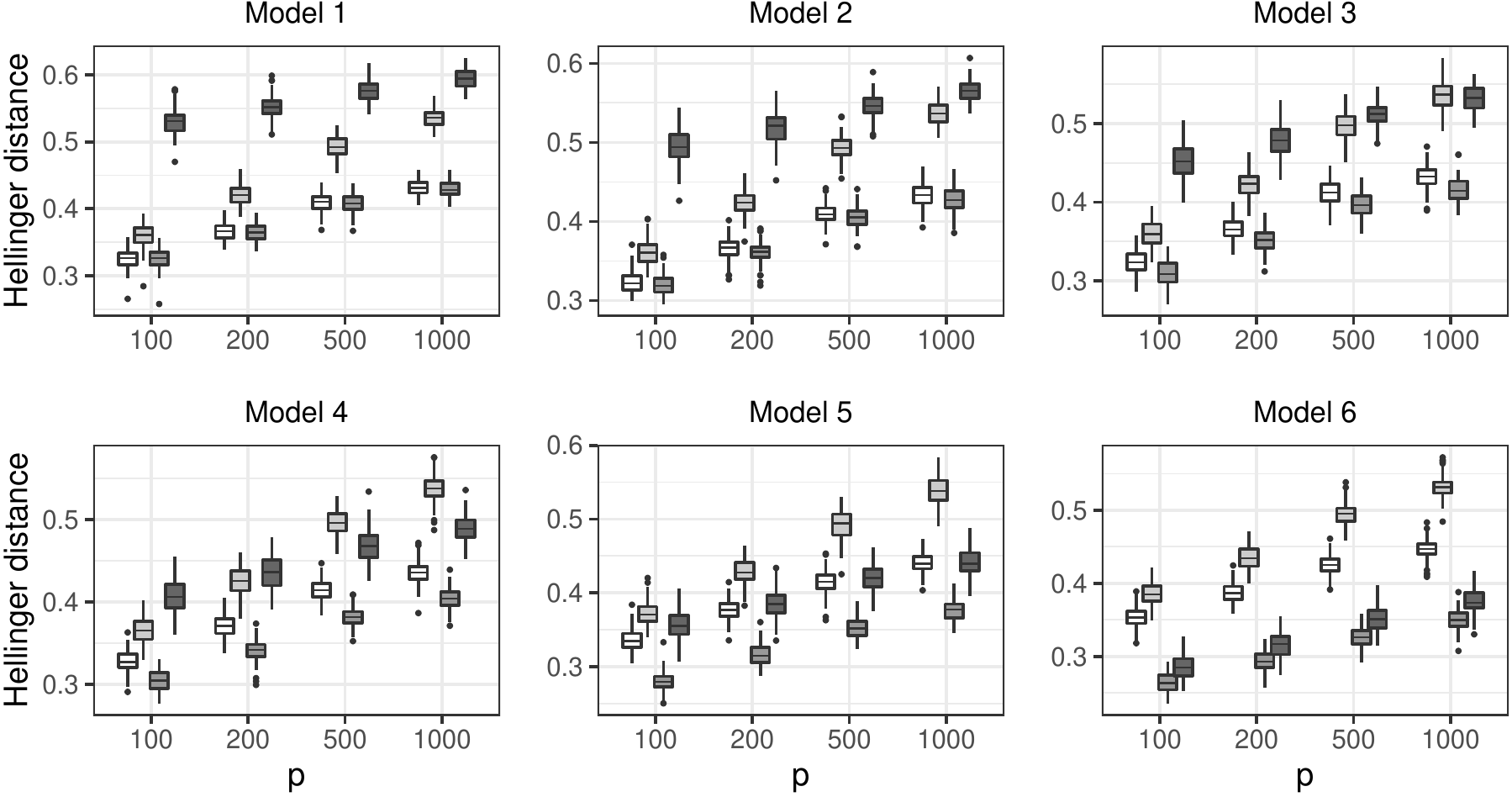}
\includegraphics[width=.5\textwidth]{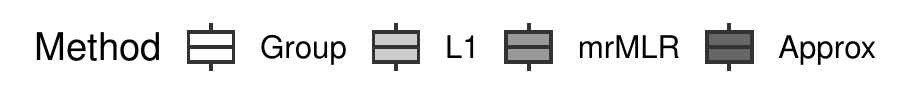}
\caption{Hellinger distance over 100 independent replications under Models 1--6 with $p \in \{100, 200, 500, 1000\}$ and $\mathcal{A}_l = \{3(l-1)+1,3(l-1)+2, 3l\}$ for $l \in \{1, 2, 3, 4\}$. }\label{fig:hellinger_NOL}
\end{figure}

In Figure \ref{fig:hellinger_NOL}, we display Hellinger distances over 100 independent replications under Models 1--6 with $p \in \{100, 200, 500, 1000\}$. Under Model 1, we see that \texttt{Group} and \texttt{mrMLR} perform almost identically. This agrees with intuition as all important predictors distinguish between all fine categories. Our method, \texttt{mrMLR}, can adapt to this situation whereas \texttt{Approx}, the only other method making use of the $\mathcal{A}_l$, performs very poorly. For Models 2--5, we see that \texttt{Group} gradually begins to perform worse relative to \texttt{mrMLR}. By Model 5 with $p = 1000$, \texttt{Approx} performs similarly to \texttt{Group}. Under Model 6, where the majority of predictors only distinguish between fine categories, \texttt{mrMLR} and \texttt{Approx} substantially outperform the other competitors.  In Wed Appendix D, we also display classification errors, Kullback-Leibler divergences, as well results on support and effect resolution recovery. In brief, the same relative performances are observed under both classification errors and Kullback-Leibler divergences. Our method tends to recover both the set of important variables and their effect resolution of $\beta^*$ with reasonable accuracy, though this is affected by both $p$ and the model. 

Beyond the competitors discussed here, we also considered ``vanilla'' random forests, hierarchical random forests \citep{kaymaz2021hierfit}, and the multiclass sparse discriminant analysis method of \citet{mai2019multiclass}. The method of \citet{kaymaz2021hierfit} was specifically designed to exploit cell type hierarchies in the context of cell type annotation. This method uses the same known coarse categories as \texttt{mrMLR}.  We display the classification accuracy of these methods versus those considered in Figure \ref{fig:hellinger_NOL} in Web Figure 8. The method of \citet{mai2019multiclass} outperforms the random forests, but none of the three methods outperform \texttt{mrMLR}. 

In Web Table 3, we provide computing times for \texttt{mrMLR}. The average time to compute our estimator for a 100 $\times$ 10 grid of candidate tuning parameters $(\gamma, \lambda)$ ranged from 6-80 minutes across the different models and predictor dimension considered in this section. 
We discuss simple ways to reduce computing times in Web Appendix D.4.

In Web Appendix D.5, we present comprehensive results for a separate simulation study with overlapping coarse categories.  Relative performances are very similar to those under the settings considered in this section.

\section{Application to interpretable cell type annotation}\label{sec:real_data}
\subsection{Overview}
We apply our method to a dataset consisting of gene expression profiles for individual peripheral blood mononuclear cells from \citet{hao2021integrated}. Immune cells within the set of peripheral blood mononuclear cells naturally exhibit a multiresolution structure, which we outline in Table \ref{tab:coarse_cell_types}. Our method is especially useful for this dataset because of the high degree of detail available in the $K=28$ cell type labels. As shown in Table \ref{tab:coarse_cell_types}, there are $L = 11$ coarse categories we consider, many of which overlap.

Our goal is to fit a model which can be used to predict the type of a cell based on its gene expression profile (measured with RNA-seq). By interpreting the coefficients estimated by our method, we may also identify genes that distinguish groups of cell types at distinct resolutions. The complete dataset (after the quality control steps described in Web Appendix F.1) consists of more than 150000 
cells from 8 individuals. 
Though protein expression data were available for these cells---and were used, in part, to assign cell type labels in \citet{hao2021integrated}---we focus on model building using only gene expression profiles because the majority of single-cell datasets do not have protein expression profiles available. Before model fitting, we normalize the gene expression profiles as described in Web Appendix F.1.

\subsection{Comparison to competing methods}\label{subsec:real_comparison}
We first compare the predictive performance of our method to a subset of competitors from Section \ref{sec:sims}. To reduce complexity, we perform (unsupervised) predictor screening on the full dataset by ranking genes as described in Web Appendix F.1 and selecting the top $p \in \{500, 750, 1000, 1500, 2000\}$ genes. For each replicate, we randomly split the dataset into training, validation, and testing sets by uniformly sampling without replacement $n \in \{10000, 20000, 30000, 40000, 50000\}$ cells for the training set and 20000 cells each for the validation and testing sets. We choose tuning parameters to minimize the validation set deviance, and measure performance using the testing set deviance and classification error rate. Additionally, we also record the degrees of freedom, defined as the number of distinct fitted parameters in the model, in order to characterize model complexity.

\begin{figure}[t]
\centering
\includegraphics[width=\textwidth]{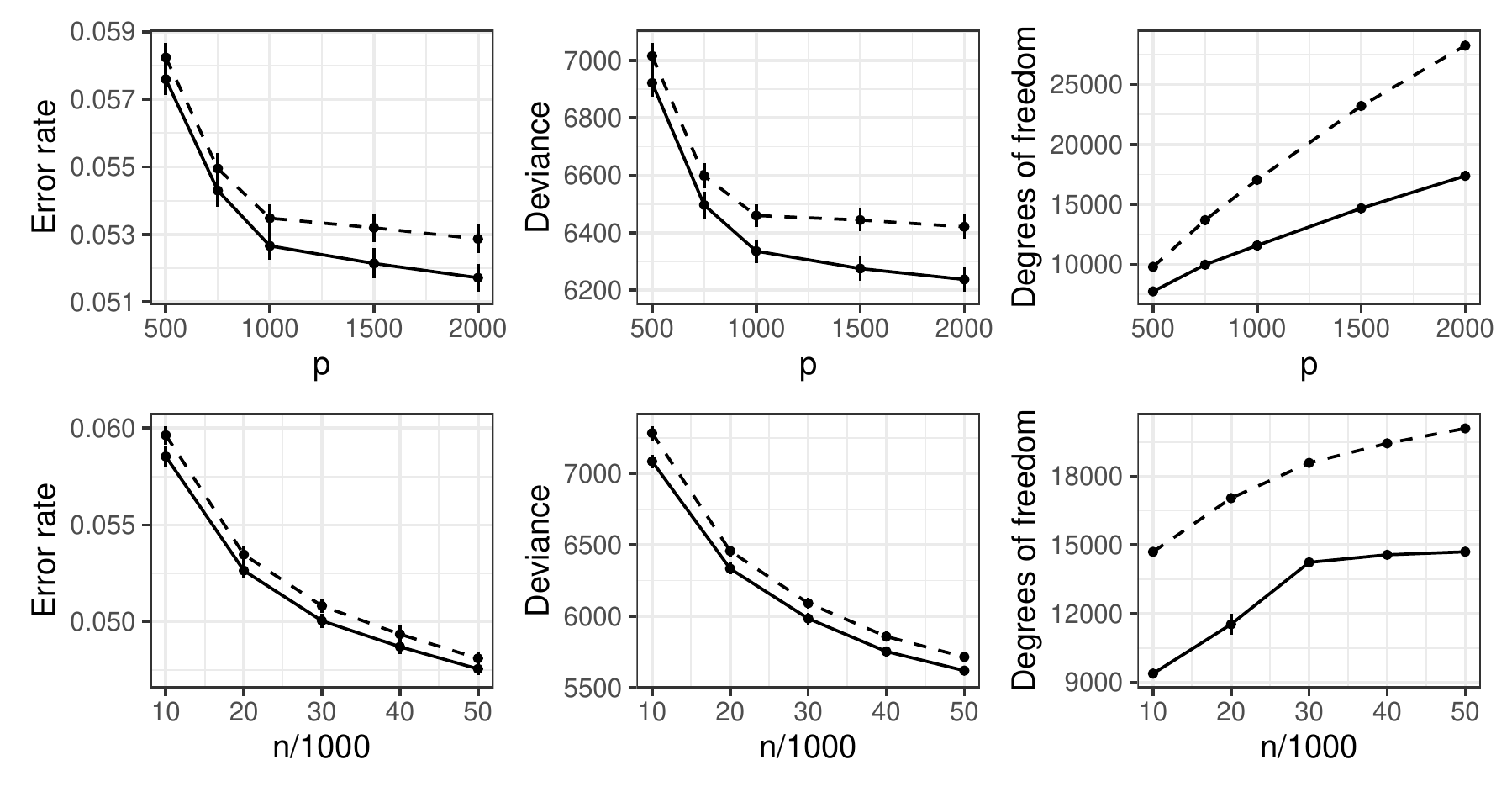}
\includegraphics[width=0.3\textwidth]{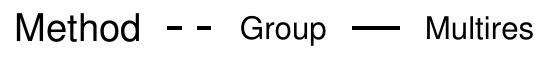}
\caption{Average classification error rates, deviances, and degrees of freedom on the single-cell RNA-seq dataset from \citet{hao2021integrated} with (top row) $n=20000$ fixed and $p \in \{500, 1000, 1500, 2000\}$ and (bottom row) $p = 1000$ fixed and $n \in \{10000, 20000, 30000, 40000, 50000\}$.}\label{fig:application_figure}
\end{figure}

We first assess the test set deviance and classification error rate when varying the number of genes $p \in \{500, 750, 1000, 1500, 2000\}$ with the training set size $n = 20000$ fixed. We repeat each setting for a total of 50 independent replications, with the training, validation, and testing sets randomly sampled in each replicate as described above. Results for \texttt{mrMLR} and \texttt{Group} are presented in the top row of Figure \ref{fig:application_figure}. For both methods, the deviance and error rate decrease with an increasing number of genes considered in the model. We observe that \texttt{mrMLR} consistently performs better than the competitors, with the gap between \texttt{mrMLR} and \texttt{Group} increasing as more genes are considered after screening. As expected, the degrees of freedom increases with larger $p$, but more gradually for \texttt{mrMLR}.  

Next, we repeat the same procedure for $n \in \{10000, 20000, 30000, 40000, 50000\}$ with the number of genes $p = 1000$ fixed. Results are presented in the bottom row of Figure \ref{fig:application_figure}. For both methods, the deviance and classification error rate decrease with an increasing training set size, with \texttt{mrMLR} again consistently outperforming \texttt{Group}. With a training set size of greater than 30000 cells, the rate of increase of the number of distinct parameters tapers off with increasing training set size for \texttt{mrMLR}, but not for \texttt{Group}.

We also compared \texttt{mrMLR} and \texttt{Group} to alternative methods---namely that of \citet{mai2019multiclass}, random forests, and hierarchical random forests \citep{kaymaz2021hierfit}. Like \texttt{mrMLR}, the method of \citet{kaymaz2021hierfit} explicitly accounts for known cell type hierarchies: we provided the method of \citet{kaymaz2021hierfit} the same hierarchy as \texttt{mrMLR} (Web Figure 15). We present results of this comparison in Web Appendix F.3. Briefly, both \texttt{mrMLR} and \texttt{Group} significantly outperform the competitors in terms of cell type classification accuracy.  

\subsection{Coherence with existing marker gene sets}\label{subsec:interp} 
\begin{figure}
\centering
~~~~~~~~~~~~~~~~~~~~~~~~\includegraphics[width=0.95\textwidth]{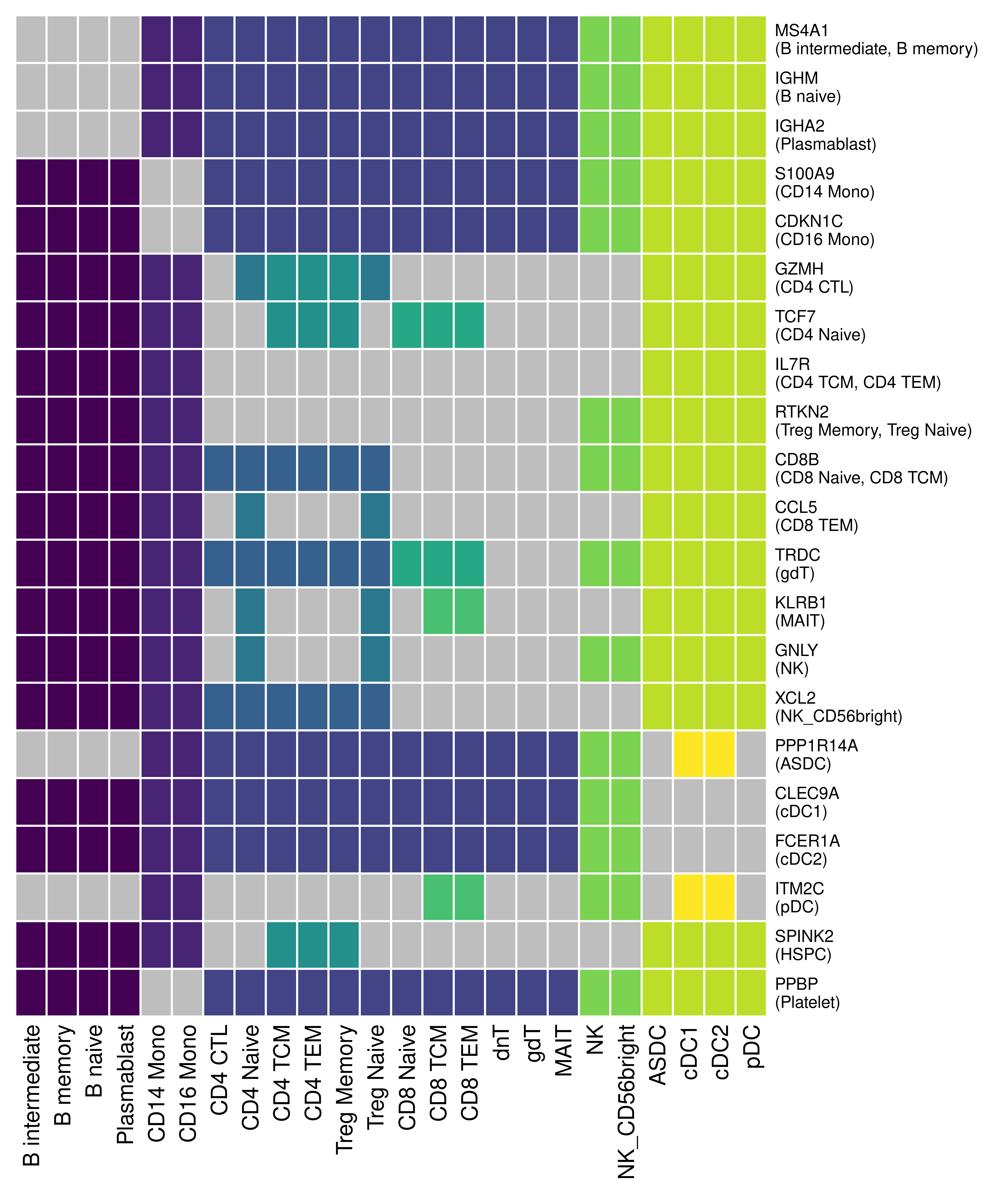}
\caption{Submatrix of $\widehat\beta$ described in Section \ref{subsec:interp}. Grey entries are those with estimated coefficient values distinct from all others in their row, whereas entries of the same (non-grey) color have identical coefficient values within a row. For example, in the first row, the coefficients corresponding to B intermediate, B memory, B naive, and Plasmablast are all distinct, whereas the coefficients are the same within subtypes of monocytes, T cells, NK cells, and dendritic cells. Vertical axis labels are genes and in parentheses, the corresponding cell type for which that gene is a ``marker gene'' in \citet{hao2021integrated}. Four fine cell types (Eryth, HSPC, ILC, Platelet) do not belong to any coarse category: their columns were omitted. 
}\label{fig:Beta_marker_gene}
\end{figure}

We focus on interpreting our estimate of the regression coefficient matrix. We fit the model using \texttt{mrMLR} with $p = 2000$ and $n = 20000$ as described in the Section \ref{subsec:real_comparison}. In Figure \ref{fig:Beta_marker_gene}, we display the estimated coefficients for genes which are most over-expressed in one cell type compared to the rest of the cell types combined (referred to as ``marker genes''), as reported in \citet{hao2021integrated}. That is, we display a submatrix of $\widehat\beta$ corresponding to the genes which are expected to be nonzero based on existing research.  In each row, we provide both the gene (e.g., in the first row, MS4A1) and the cell types in which this gene has been reported to be the most over-expressed gene (e.g., in the first row, MS4A1 is over-expressed in both B intermediate and B memory cells). A version of Figure \ref{fig:Beta_marker_gene} for the entire regression coefficient matrix estimate can be found in Web Figure 6.  In Web Figure 13 and 14, we provide plots of the expression of GZMH, MS4A1, and XCL2---three genes in Figure \ref{fig:Beta_marker_gene}---partitioned over cell types they are estimated to distinguish between. 

For the sake of example, we focus on the first three rows in Figure \ref{fig:Beta_marker_gene}. These rows contain the estimated coefficients for genes previously reported to be over-expressed in subtypes of B cells. We see that the coefficients for these genes are distinct for B cell subtype categories (B intermediate, B memory, B naive, Plasmablast), but are the same within subtypes of monocytes, T cells, NK cells, and dendritic cells. Hence, we would interpret MS4A1, IGHM, and IGHA2 as useful for distinguishing between subtypes of B cells, and for discriminating between B cells, monocytes, T cells, NK cells, and dendritic cells, but not useful for discriminating between subtypes of monocytes, T cells, NK cells, or dendritic cells.
This interpretation coheres with the assignment of these genes as marker genes for B cell subtypes in \citet{hao2021integrated}. We did not use this marker gene information in model fitting.  

Some of our results are arguably more informative than the marker gene assignments from \citet{hao2021integrated}. For example, \citet{hao2021integrated} label GZMH as a marker gene for CD4$^+$ CTL, whereas our method estimates GZMH to distinguish between all T cell subtypes except subtypes of CD4$^+$ naive and CD4$^+$ memory T cells. Such a difference in interpretation follows from the fact that our model-based method accounts for the joint behavior of gene expression in distinguishing between cell types. The published marker genes, in contrast, were identified via univariate differential expression testing after cell type annotation.

To verify that our set of estimated marker genes is competitive with existing marker gene sets, we repeated the analyses from Section \ref{subsec:real_comparison} wherein we used only the marker genes from \citet{hao2021integrated} as candidate predictors for both \texttt{mrMLR} and \texttt{RF}. Details are provided in Web Appendix F.3. The version of our method which selects genes from a large candidate set performs better than both the methods which use the marker genes from \citet{hao2021integrated} in terms of classification accuracy. These analyses demonstrate that our method could be used in conjunction with existing methods for identifying marker genes. For example, one could use the method of \citet{dumitrascu2021optimal} to select marker genes, then input only these genes into \texttt{mrMLR} with $\gamma = 0$ so that all marker genes are retained by the model.   

\section{Discussion}
This work is focused on point estimation for \eqref{eq:multinomiaL_model} for the sake of classification and coefficient interpretation. A natural question is how to test the structures (sparsity pattern and effect resolution) discovered by our method. Fortunately, due to the relatively large sample sizes in modern single-cell RNA-seq datasets, one can perform likelihood-based inference in a straightforward way using sample splitting. This would proceed in three steps:  (i) randomly partition cells into two sets: training and inference; (ii) using \texttt{mrMLR}, fit \eqref{eq:multinomiaL_model} to the training data, performing cross-validation to select tuning parameters; (iii) test the selected model's sparsity and effect resolution by fitting the constrained MLE on the inference set and applying standard inferential techniques on the fitted model. Although sample splitting-based inference can come at the cost of decreased power, in our real data analysis $n$ is quite large ($n > 150000$), so we expect sample splitting to provide reasonable power. \\

\noindent \textbf{Acknowledgements.}
The authors thank the associate editor and referee for their helpful comments and suggestions. 
A. J. Molstad's research was supported by a grant from the National Science Foundation (DMS-2113589). \\
\smallskip

\noindent \textbf{Data availability statement.}
The data that support the findings in this paper are available through the R package \texttt{AnnotatedPBMC}, available at \url{https://github.com/keshav-motwani/AnnotatedPBMC}. 

 \bibliographystyle{biom} 

\bibliography{HierMultinom}

\begin{thebibliography}{}

\bibitem[\protect\citeauthoryear{Agresti}{Agresti}{2003}]{agresti2003categorical}
Agresti, A. (2003).
\newblock {\em Categorical data analysis}.
\newblock John Wiley and Sons.

\bibitem[\protect\citeauthoryear{Beck and Teboulle}{Beck and
  Teboulle}{2009}]{beck2009fast}
Beck, A. and Teboulle, M. (2009).
\newblock A fast iterative shrinkage-thresholding algorithm for linear inverse
  problems.
\newblock {\em SIAM Journal on Imaging Sciences} {\bf 2,} 183--202.

\bibitem[\protect\citeauthoryear{Bernstein, Ma, Gleicher, and Dewey}{Bernstein
  et~al.}{2021}]{bernstein2021cello}
Bernstein, M.~N., Ma, Z., Gleicher, M., and Dewey, C.~N. (2021).
\newblock {CellO}: Comprehensive and hierarchical cell type classification of
  human cells with the cell ontology.
\newblock {\em Iscience} {\bf 24,}.

\bibitem[\protect\citeauthoryear{de~Kanter, Lijnzaad, Candelli, Margaritis, and
  Holstege}{de~Kanter et~al.}{2019}]{de2019chetah}
de~Kanter, J.~K., Lijnzaad, P., Candelli, T., Margaritis, T., and Holstege,
  F.~C. (2019).
\newblock {CHETAH}: a selective, hierarchical cell type identification method
  for single-cell {RNA} sequencing.
\newblock {\em Nucleic Acids Research} {\bf 47,} e95--e95.

\bibitem[\protect\citeauthoryear{Dumitrascu, Villar, Mixon, and
  Engelhardt}{Dumitrascu et~al.}{2021}]{dumitrascu2021optimal}
Dumitrascu, B., Villar, S., Mixon, D.~G., and Engelhardt, B.~E. (2021).
\newblock Optimal marker gene selection for cell type discrimination in single
  cell analyses.
\newblock {\em Nature Communications} {\bf 12,} 1--8.

\bibitem[\protect\citeauthoryear{Hao, Hao, Andersen-Nissen, Mauck~III, Zheng,
  Butler, Lee, Wilk, Darby, and Zager}{Hao et~al.}{2021}]{hao2021integrated}
Hao, Y., Hao, S., Andersen-Nissen, E., Mauck~III, W.~M., Zheng, S., Butler, A.,
  Lee, M.~J., Wilk, A.~J., Darby, C., and Zager, M. (2021).
\newblock Integrated analysis of multimodal single-cell data.
\newblock {\em Cell} {\bf 184,} 3573--3587.

\bibitem[\protect\citeauthoryear{Kaymaz, Ganglberger, Tang, Haslinger,
  Fernandez-Albert, Lawless, and Sackton}{Kaymaz
  et~al.}{2021}]{kaymaz2021hierfit}
Kaymaz, Y., Ganglberger, F., Tang, M., Haslinger, C., Fernandez-Albert, F.,
  Lawless, N., and Sackton, T.~B. (2021).
\newblock Hierfit: a hierarchical cell type classification tool for projections
  from complex single-cell atlas datasets.
\newblock {\em Bioinformatics} {\bf 37,} 4431--4436.

\bibitem[\protect\citeauthoryear{L{\"a}hnemann, K{\"o}ster, Szczurek, McCarthy,
  Hicks, Robinson, Vallejos, Campbell, Beerenwinkel, and Mahfouz}{L{\"a}hnemann
  et~al.}{2020}]{lahnemann2020eleven}
L{\"a}hnemann, D., K{\"o}ster, J., Szczurek, E., McCarthy, D.~J., Hicks, S.~C.,
  Robinson, M.~D., Vallejos, C.~A., Campbell, K.~R., Beerenwinkel, N., and
  Mahfouz, A. (2020).
\newblock Eleven grand challenges in single-cell data science.
\newblock {\em Genome Biology} {\bf 21,} 1--35.

\bibitem[\protect\citeauthoryear{Maecker, McCoy, and Nussenblatt}{Maecker
  et~al.}{2012}]{maecker2012standardizing}
Maecker, H.~T., McCoy, J.~P., and Nussenblatt, R. (2012).
\newblock Standardizing immunophenotyping for the human immunology project.
\newblock {\em Nature Reviews Immunology} {\bf 12,} 191--200.

\bibitem[\protect\citeauthoryear{Mai, Yang, and Zou}{Mai
  et~al.}{2019}]{mai2019multiclass}
Mai, Q., Yang, Y., and Zou, H. (2019).
\newblock Multiclass sparse discriminant analysis.
\newblock {\em Statistica Sinica} {\bf 29,} 97--111.

\bibitem[\protect\citeauthoryear{Molstad and Rothman}{Molstad and
  Rothman}{2023}]{Molstad2021}
Molstad, A.~J. and Rothman, A.~J. (2023).
\newblock A likelihood-based approach for multivariate categorical response
  regression in high dimensions.
\newblock {\em Journal of the American Statistical Association} {\bf 118,}
  1402--1414.

\bibitem[\protect\citeauthoryear{Motwani, Bacher, and Molstad}{Motwani
  et~al.}{2023}]{motwani2021binned}
Motwani, K., Bacher, R., and Molstad, A.~J. (2023).
\newblock Binned multinomial logistic regression for integrative cell type
  annotation.
\newblock {\em Annals of Applied Statistics} .

\bibitem[\protect\citeauthoryear{Negahban, Ravikumar, Wainwright, and
  Yu}{Negahban et~al.}{2012}]{negahban2012unified}
Negahban, S.~N., Ravikumar, P., Wainwright, M.~J., and Yu, B. (2012).
\newblock A unified framework for high-dimensional analysis of m-estimators
  with decomposable regularizers.
\newblock {\em Statistical Science} {\bf 27,} 538--557.

\bibitem[\protect\citeauthoryear{Nibbering and Hastie}{Nibbering and
  Hastie}{2022}]{nibbering2022multiclass}
Nibbering, D. and Hastie, T.~J. (2022).
\newblock Multiclass-penalized logistic regression.
\newblock {\em Computational Statistics and Data Analysis} {\bf 169,} 107414.

\bibitem[\protect\citeauthoryear{Pasquini, Arias, Sch{\"a}fer, and
  Busskamp}{Pasquini et~al.}{2021}]{pasquini2021automated}
Pasquini, G., Arias, J. E.~R., Sch{\"a}fer, P., and Busskamp, V. (2021).
\newblock Automated methods for cell type annotation on sc{RNA}-seq data.
\newblock {\em Computational and Structural Biotechnology Journal} .

\bibitem[\protect\citeauthoryear{Polson, Scott, and Willard}{Polson
  et~al.}{2015}]{polson2015proximal}
Polson, N.~G., Scott, J.~G., and Willard, B.~T. (2015).
\newblock Proximal algorithms in statistics and machine learning.
\newblock {\em Statistical Science} {\bf 30,} 559--581.

\bibitem[\protect\citeauthoryear{Powers, Hastie, and Tibshirani}{Powers
  et~al.}{2018}]{powers2018nuclear}
Powers, S., Hastie, T., and Tibshirani, R. (2018).
\newblock Nuclear penalized multinomial regression with an application to
  predicting at bat outcomes in baseball.
\newblock {\em Statistical Modelling} {\bf 18,} 388--410.

\bibitem[\protect\citeauthoryear{Price, Geyer, and Rothman}{Price
  et~al.}{2019}]{price2019automatic}
Price, B.~S., Geyer, C.~J., and Rothman, A.~J. (2019).
\newblock Automatic response category combination in multinomial logistic
  regression.
\newblock {\em Journal of Computational and Graphical Statistics} {\bf 28,}
  758--766.

\bibitem[\protect\citeauthoryear{Vincent and Hansen}{Vincent and
  Hansen}{2014}]{vincent2014sparse}
Vincent, M. and Hansen, N.~R. (2014).
\newblock Sparse group lasso and high dimensional multinomial classification.
\newblock {\em Computational Statistics and Data Analysis} {\bf 71,} 771--786.

\bibitem[\protect\citeauthoryear{Yan and Bien}{Yan and
  Bien}{2017}]{yan2017hierarchical}
Yan, X. and Bien, J. (2017).
\newblock Hierarchical sparse modeling: A choice of two group lasso
  formulations.
\newblock {\em Statistical Science} {\bf 32,} 531--560.

\bibitem[\protect\citeauthoryear{Yan and Bien}{Yan and
  Bien}{2021}]{yan2021rare}
Yan, X. and Bien, J. (2021).
\newblock Rare feature selection in high dimensions.
\newblock {\em Journal of the American Statistical Association} {\bf 116,}
  887--900.

\bibitem[\protect\citeauthoryear{Yee and Hastie}{Yee and
  Hastie}{2003}]{yee2003reduced}
Yee, T.~W. and Hastie, T.~J. (2003).
\newblock Reduced-rank vector generalized linear models.
\newblock {\em Statistical Modelling} {\bf 3,} 15--41.

\bibitem[\protect\citeauthoryear{Yuan and Lin}{Yuan and
  Lin}{2006}]{yuan2006model}
Yuan, M. and Lin, Y. (2006).
\newblock Model selection and estimation in regression with grouped variables.
\newblock {\em Journal of the Royal Statistical Society: Series B} {\bf 68,}
  49--67.

\bibitem[\protect\citeauthoryear{Zhu and Hastie}{Zhu and
  Hastie}{2004}]{zhu2004classification}
Zhu, J. and Hastie, T. (2004).
\newblock Classification of gene microarrays by penalized logistic regression.
\newblock {\em Biostatistics} {\bf 5,} 427--443.

\end{thebibliography}
\section*{Supporting Information}
Web Appendices, Tables, and Figures, referenced in Sections 2--6 are available with this paper at the Biometrics website on Wiley Online Library. Data and code needed to reproduce the results from Sections 6 and 7, as well as an R package implementing \texttt{mrMLR}, is available at \texttt{https://github.com/ajmolstad/HierMultinom}.
\end{document}